\documentclass[aps,prl,twocolumn,showpacs]{revtex4-1}
\usepackage{graphicx}  
\usepackage{subfigure}  
\usepackage{bm}        
\usepackage{amssymb}   
\usepackage[tbtags]{amsmath}
\usepackage{exscale}
\usepackage{xcolor}
\newcommand{\ket}[1]{|#1\rangle}
\newcommand{\bra}[1]{\langle#1|}

\definecolor{darkred}{rgb}{0.90,0.2,0.2}

\usepackage[plainpages=false,pdfpagelabels,colorlinks=true,linkcolor=red,urlcolor=blue,citecolor=blue,pdftitle={Heating rates of strongly interacting quantum systems},pdfauthor={},pdfdisplaydoctitle=true,pdfduplex=DuplexFlipLongEdge]{hyperref}

\begin{document}

\title{Heating Rates in Periodically Driven Strongly Interacting\\ Quantum Many-Body Systems}

\author{Krishnanand Mallayya}
\affiliation{Department of Physics, The Pennsylvania State University, University Park, Pennsylvania 16802, USA}
\author{Marcos Rigol}
\affiliation{Department of Physics, The Pennsylvania State University, University Park, Pennsylvania 16802, USA}


\begin{abstract}
We study heating rates in strongly interacting quantum lattice systems in the thermodynamic limit. Using a numerical linked cluster expansion, we calculate the energy as a function of the driving time and find a robust exponential regime. The heating rates are shown to be in excellent agreement with Fermi's golden rule. We discuss the relationship between heating rates and, within the eigenstate thermalization hypothesis, the smooth function that characterizes the off-diagonal matrix elements of the drive operator in the eigenbasis of the static Hamiltonian. We show that such a function, in nonintegrable and (remarkably) integrable Hamiltonians, can be probed experimentally by studying heating rates as functions of the drive frequency.  
\end{abstract}

\maketitle

Periodic perturbations are a ubiquitous tool to excite and probe quantum systems and study their response functions. Recent developments in theory and experiments have expanded the scope of periodic driving to generate effective magnetic fields \cite{Dalibard_2011, Kitagawa_2011, Aidelsburger_2013, Goldman_2014}, as well as to engineer topologically nontrivial band structures \cite{oka_aoki_09, Lindner2011, Rechtsman2013, Cooper_2019} and novel time-crystalline phases \cite{Else_2016_Floquet, Else_2017_Prethermal, Khemani_2016_Phase, Yao_2017_Discrete, *Yao_2017_Erratum, zhang2017observation, choi2017observation}. However, under periodic driving, generic many-body interacting systems are expected to heat up and (for a bounded spectrum, typical of lattice fermions and spins) equilibrate at long times to states that are effectively at infinite temperature~\cite{DAlessio_2014, Lazarides_2014}. 

Driving at high frequencies, because of prethermalization~\cite{moeckel_kehrein_2008, *moeckel_kehrein_2009, eckstein_kollar_09, kollar_wolf_11, tavora_mitra_13, tavora_rosch_14, nessi_iucci_14, essler2014quench, bertini2015prethermalization, *bertini2016prethermalization, canovi2016stroboscopic, fagotti2015universal, lange2018time, lenarcic2018perturbative, reimann_dabelow_19, Mallayya_2019_Prethermalization}, has been proposed to slow down heating~\cite{Abanin_2015_Exponentially, Else_2017_Prethermal, machado2017exponentially, weidinger2017floquet, kuhlenkamp2019periodically}. It results in initial fast prethermal dynamics towards time-periodic steady states (prethermal states) of effective local Hamiltonians~\cite{Abanin_2017_Effective, abanin2017rigorous, Mori_2016_Rigorous, Kuwahara_2016_Floquet}, before thermalization dynamics eventually results in featureless ``infinite-temperature'' states~\cite{DAlessio_2014, Lazarides_2014, Prosen_1998, DAlessio_2013}. Prethermalization is a universal phenomenon that occurs during dynamics in isolated~\cite{Mallayya_2019_Prethermalization} and open~\cite{lange2018time,lenarcic2018perturbative} systems whenever conservation laws are weakly broken. Numerical studies of prethermalization and thermalization, or, in general, of energy absorption in driven strongly interacting systems with many particles (or spins) are challenging. Progress has been achieved using massively parallel Krylov subspace methods~\cite{machado2017exponentially}, density matrix truncation~\cite{ye2019emergent}, and t-DMRG~\cite{kollath2016}, but there is a dearth of computational techniques to study generic models in arbitrary dimensions.

Here, we report on the implementation of a numerical linked cluster expansion (NLCE) for driven systems. NLCEs can be used to study arbitrary interaction strengths in arbitrary dimensions. They were originally introduced to study thermal equilibrium ensembles \cite{rigol2006numerical, *rigol2007numerical, *rigol2007numerical2}, where they outperform full exact diagonalization calculations~\cite{iyer_srednicki_15}. NLCEs were recently implemented to study thermalization \cite{rigol_14a, *rigol_16} and quantum dynamics under time-independent Hamiltonians in one~\cite{Mallayya_2018_Quantum, Mallayya_2019_Prethermalization} and two~\cite{white2017quantum, Guardado_2018_Probing} dimensions, and combined with dynamical quantum typicality \cite{Richter_2019_Combining}. We use them to determine heating rates in strongly interacting one-dimensional (1D) lattices in the thermodynamic limit. The numerically obtained rates are shown to agree with Fermi's golden rule predictions. We argue that, in addition to helping quantify the stability of prethermal states, heating rates can be used to probe the structure of the off-diagonal matrix elements of the drive operator in the eigenstates of the static Hamiltonian.

We consider a time-periodic Hamiltonian of the form $\hat{H}(\tau)=\hat{H}_0+g(\tau)\hat{K}$, where $\hat{H}_0$ is the static Hamiltonian and $g(\tau)\hat{K}$ is a weak time-periodic perturbation of strength $g$, period $T=2\pi/\Omega$, and zero time average. The system is initialized (at $\tau=0$) in a state $\hat{\rho}_I=\exp[-\beta_I\hat{H}_I]/\text{Tr}\{\exp[-\beta_I\hat{H}_I]\}$ that is a thermal equilibrium state of an initial static Hamiltonian $\hat{H}_I$ at an inverse temperature $\beta_I$.  At stroboscopic times $\tau=nT$  $(n=0,1,2,\dots)$, the density matrix $\hat{\rho}(\tau)$ can be written as $\hat{\rho}(\tau)=(\hat{U}_F)^n\hat{\rho}_I(\hat{U}_F^{\dagger})^n$, where $\hat{U}_F=\mathcal{T}\exp[-i\int_{0}^{T}\hat{H}(t)dt]$ is the (time ordered $\mathcal{T}$) Floquet evolution operator (we set $\hbar=1$). We assume that $\hat{H}_I$, $\hat{H}_0$, and $\hat{K}$ are translationally invariant sums of local operators, and that they are mutually noncommuting (nontrivial dynamics occurs even if $g=0$). 

The obvious conservation law broken by $g(\tau)\hat{K}$ is energy conservation. For sufficiently small $g$ in the thermodynamic limit, we expect prethermalization to occur (independently of the value of $\Omega$), wherein the system quickly relaxes to the equilibrium state of $\hat{H}_0$ described by a (generalized) Gibbs ensemble [up to $O(g)$ corrections]. The relaxation towards infinite temperature can be described by a slowly evolving (generalized) Gibbs ensemble of $\hat{H}_0$, characterized by the instantaneous expectation values of the conserved quantities of $\hat{H}_0$~\cite{Mallayya_2019_Prethermalization}. The dynamics of those quantities is described by autonomous equations, with drifts given by Fermi's golden rule~\cite{Mallayya_2019_Prethermalization}. 

We study the evolution of the energy defined by the static Hamiltonian, which is also the time-averaged Hamiltonian $\overline{\hat{H}(\tau)}=\hat{H}_0$, $E(\tau)=\text{Tr}[\hat{H}_0\hat{\rho}(\tau)]$. We consider general time-periodic perturbations, which can be Fourier decomposed as $g(\tau)\hat{K}=\sum_{m>0} 2g_m \sin(m\Omega\tau)\hat{K}$. After a short initial transient dynamics, in the linear response regime, the system absorbs energy independently from each Fourier mode $m$. The average rate of energy absorption over a cycle is $\dot{E}(\tau)=\sum_{m>0}\dot{E}_m(\tau)$ with, as expected from Fermi's golden rule,
\begin{eqnarray}
\dot{E}_m(\tau)=2\pi g_m^2\sum_{\substack{i,f}}&&|\bra{E^0_f}\hat{K}\ket{E^0_i}|^2(E^0_f-E^0_i)P_i^0(\tau)\nonumber\\
&&\times\delta(E^0_f-E^0_i\pm m\Omega) \label{FGR_eq},
\end{eqnarray}  
where $\ket{E^0_{i}}$ ($\ket{E^0_{f}}$) are the eigenkets of $\hat{H}_0$ with eigenenergies $E^0_{i}$ ($E^0_{f}$), and $P_i^0(\tau)=\bra{E^0_i}\hat{\rho}(\tau)\ket{E^0_i}$ is the projection of $\hat{\rho}(\tau)$ into the basis of $\hat{H}_0$. The latter defines the so-called diagonal ensemble (DE) at time $\tau$~\cite{rigol2008thermalization}, $\hat{\rho}_{\text{DE}}(\tau)=P_i^0(\tau)\ket{E_i^0}\bra{E_i^0}$. $\hat{\rho}_{\text{DE}}(\tau)$ is expected to characterize the equilibrated state under $\hat{H}_0$ at time $\tau$~\cite{dalessio_kafri_16}. We define the rate $\Gamma(\tau)=\sum_{m>0}\Gamma_m(\tau)$, where $\Gamma_m(\tau)=\dot{E}_m(\tau)/[E_\infty-E(\tau)]$ is the rate for Fourier mode $m$, and $E_\infty$ is the energy at infinite temperature. Only when it is sufficiently small does one expect $|E_\infty-E(\tau)|$ to be an exponential function, and $\Gamma(\tau)$ to be meaningful.

We focus on 1D lattice system of hard-core bosons, with $\hat{H}_0$ and $\hat{K}$ given by
\begin{eqnarray}
&&\hat{H}_0=\sum_i \left[ \left(
-t\, \hat{b}^\dagger_i \hat{b}^{}_{i+1} - t'\, \hat{b}^\dagger_i \hat{b}^{}_{i+2} +h\, \hat{b}^\dagger_i\right)  + \text{H.c.} \right. \label{Ham}\\
&&\left.+V\left(\hat{n}^{}_i-\dfrac{1}{2}\right)\hspace*{-0.1cm}\left(\hat{n}^{}_{i+1}-\dfrac{1}{2}\right) +V'\left(\hat{n}^{}_i-\dfrac{1}{2}\right)\hspace*{-0.1cm}\left(\hat{n}^{}_{i+2}-\dfrac{1}{2}\right)\right],\nonumber\\&&\hat{K}=-\sum_i \left( \hat{b}^\dagger_i \hat{b}^{}_{i+1} + \text{H.c.} \right),
\end{eqnarray}
where standard notation was used~\cite{cazalilla2011one}. We drive the system with a square wave $g(\tau)=g\, \text{sgn}[\sin(\Omega\tau)]$, and set $t=V=1$ (our unit of energy and frequency). $\hat{H}_0$ is integrable for $t'=V'=h=0$ (and mappable to the spin-1/2 $XXZ$ Hamiltonian \cite{cazalilla2011one}), and nonintegrable for nonvanishing $t'$, $V'$, and $h$. We study integrable and nonintegrable (with $t'=V'=0.8$ and $h=1.0$) cases, and select $\hat{H}_I$ to have the same terms as  $\hat{H}_0$ [Eq.~\eqref{Ham}] but with different nearest neighbor coupling parameters ($t_I=0.5$ and $V_I=2.0$).

\begin{figure}[!t]
\includegraphics[width=0.475\textwidth]{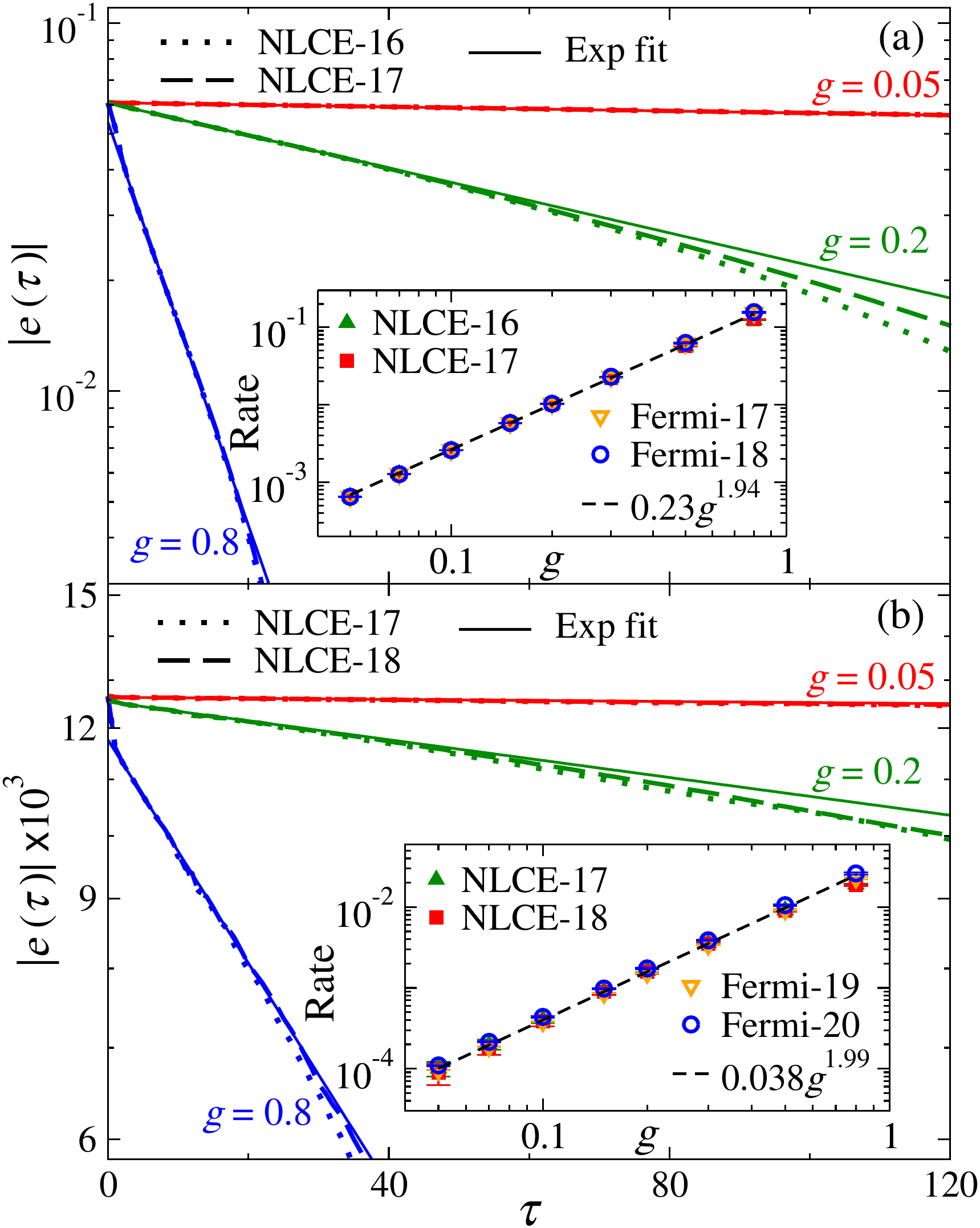}
\vspace{-0.05cm}
\caption{(Main panels) Absolute value of the energy per site $|e(\tau)|$ vs $\tau$ for (a) the nonintegrable and (b) the integrable $\hat{H}_0$ for three strengths $g=\{0.05,0.2,0.8\}$ of the drive, a period $T=1.0$, and $\beta_I=(30)^{-1}$. Results (at stroboscopic times) are obtained using NLCE to (a) 16 (NLCE-16) and 17 (NLCE-17) orders, and (b) 17 (NLCE-17) and 18 (NLCE-18) orders. The solid lines show exponential fits to the highest NLCE order. (Insets) Rates obtained in fits, as those depicted in the main panels, for the two highest NLCE orders. For all values of $g$, the fits for the nonintegrable $\hat{H}_0$ are done for times $3\le\tau\le20$ for NLCE-17 and $3\le\tau\le15$ for NLCE-16, while for the integrable $\hat{H}_0$ they are done for times $2\le\tau\le8$ for NLCE-18 and $2\le\tau\le7$ for NLCE-17. The Fermi golden rule predictions (open symbols) are evaluated using full exact diagonalization in chains with: (a) 17 and 18 sites (Fermi-17 and Fermi-18) and (b) 19 and 20 sites (Fermi-19 and Fermi-20), and periodic boundary conditions. Error bars indicate the fitting errors for the NLCE rates, and the standard deviation from averages over different values of $\Delta E$ and $\tau$ for the Fermi golden rule predictions~\cite{suppmat}. Power-law fits ($\alpha g^{\gamma}$) of the rates in both insets are done for the highest order of the NLCE in the interval $0.05 \le g \le 0.3$.}\label{Fig_rate_g}
\end{figure}

We implement a NLCE to calculate the energy per site $e(\tau)=E(\tau)/L$ at stroboscopic times in the thermodynamic limit $(L\rightarrow\infty)$. Within NLCEs, $e(\tau)$ is expressed as a sum over the contributions of all connected clusters ($c$) that can be embedded on the lattice, $e(\tau)=\sum_c M(c)\times W^e_c(\tau)$, where $M(c)$ is the number of ``embeddings'' (per site) of cluster $c$, and $W^e_c(\tau)$ is the weight of $e(\tau)$ in cluster $c$. $W^e_c(\tau)$ is obtained recursively using the inclusion-exclusion principle: $W^e_c(\tau)=E_c(\tau)-\sum_{c'\subset c}W^e_{c'}(\tau)$, where $c'$ denotes the connected subclusters of $c$ and $E_c(\tau)=\text{Tr}[\hat{H}_0^c\hat{\rho}_c(\tau)]$ is the energy in cluster $c$ [$\hat{H}^c_0$ is the static Hamiltonian, and $\hat{\rho}_c(\tau)$ is the density matrix at time $\tau$, both in cluster $c$]. The series starts with the smallest cluster (a site) for which $W_c(\tau)=E_c(\tau)$. For each cluster, $E_c(\tau)$ is calculated numerically using full exact diagonalization. We use maximally connected clusters (clusters with contiguous sites and all possible bonds) as they are optimal to study dynamics in chains in the presence of nearest and next-nearest neighbor interactions~\cite{rigol_14a, *rigol_16, Mallayya_2018_Quantum, Mallayya_2017_Numerical}. The order of the NLCE is set by the number of sites of the largest cluster considered. For nonintegrable $\hat{H}_0$, we compute 17 orders of the NLCE (after exploiting all symmetries, the dimension of largest sector of the Hamiltonian is 32\,896). When $\hat{H}_0$ is integrable, due to particle number conservation, we are able to compute 18 orders of the NLCE (the dimension of the largest sector in this case is 21\,942).

In the main panels of Fig.~\ref{Fig_rate_g}, we show NLCE results for $|e(\tau)|$ vs $\tau$ for (a) the nonintegrable and (b) the integrable static Hamiltonians, for three strengths $g=0.05$, 0.2, and 0.8 of the drive, for an initial thermal equilibrium state of $\hat{H}_I$ at an inverse temperature $\beta_I=(30)^{-1}$. The exponential fits, which exclude the short-time transient dynamics and long times at which the NLCE does not converge, make apparent that the approach of $e(\tau)$ to the infinite-temperature energy ($E_{\infty}/L=0$) is exponential. The rates obtained from such fits are plotted in the insets of Fig.~\ref{Fig_rate_g} vs $g$, for the two highest orders of the NLCE. They agree with each other, indicating that the fits are robust. The rates are $\propto g^2$ and are in excellent agreement with Fermi's golden rule [Eq.~\eqref{FGR_eq}], evaluated numerically using full exact diagonalization in chains with periodic boundary conditions~\cite{suppmat}.

\begin{figure}[!t]
\includegraphics[width=0.475\textwidth]{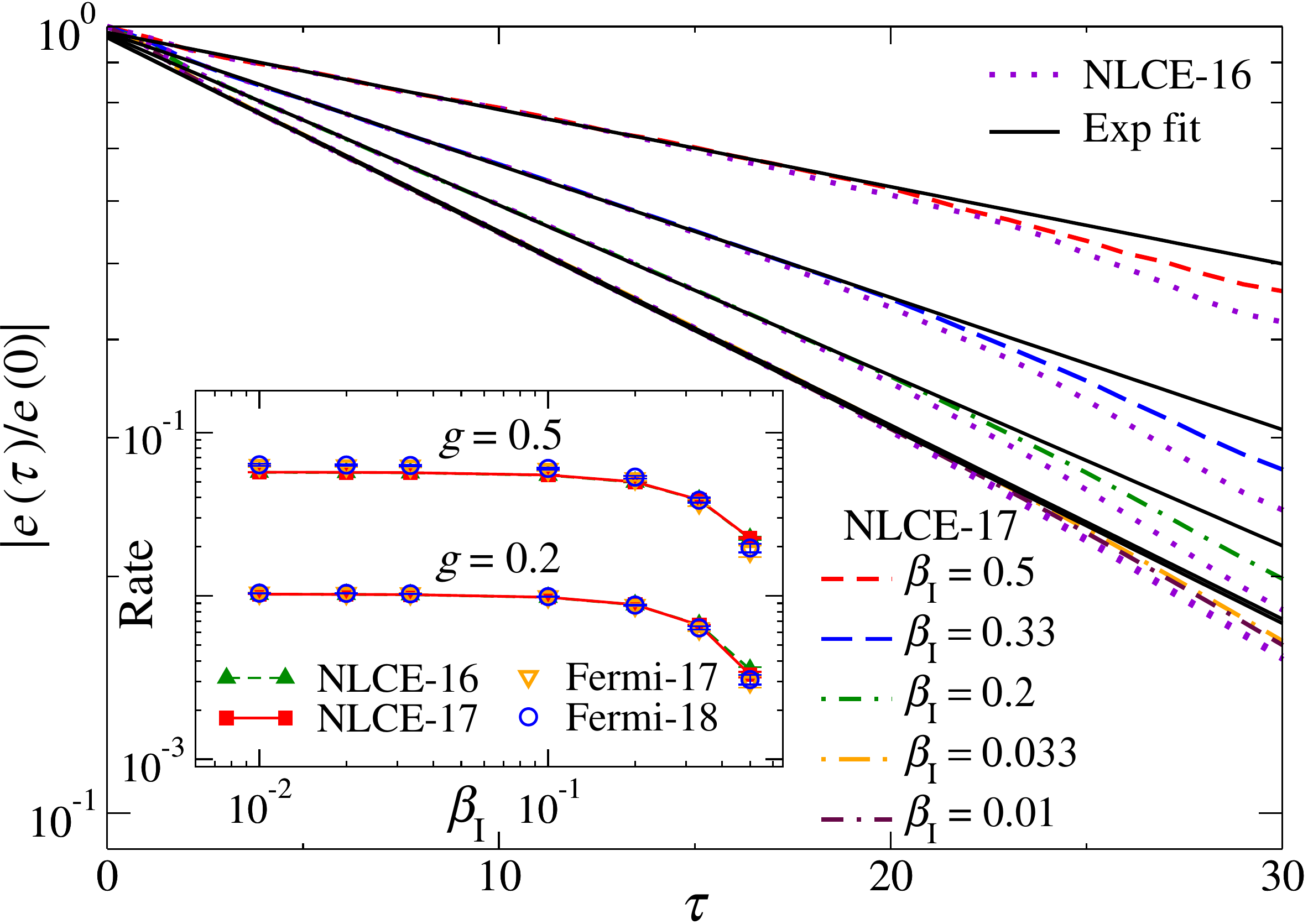}
\vspace{-0.05cm}
\caption{(Main panel) Absolute value of the energy per site $|e(\tau)|$, normalized by its initial value $|e(0)|$, for a periodically driven nonintegrable $\hat{H}_0$ with $g=0.5$ and $T=1.0$, for initial thermal states of $\hat{H}_I$ at different inverse temperatures $\beta_I$. We show results for 16 and 17 orders of the NLCE (NLCE-16 and NLCE-17, respectively), and exponential fits to the NLCE-17 results. (Inset) Rates obtained from exponential fits to NLCE-17 for $3\le\tau\le20$ (as those in the main panel) and NLCE-16 for $3\le\tau\le15$ vs $\beta_I$, for $g=0.2$ and $g=0.5$. We also report Fermi's golden rule predictions obtained using full exact diagonalization in chains with 17 and 18 sites (Fermi-17 and Fermi-18) and periodic boundary conditions.}\label{Fig_rate_beta}
\end{figure}

It follows from eigenstate thermalization for nonintegrable Hamiltonians~\cite{deutsch1991quantum, srednicki1994chaos, rigol2008thermalization, dalessio_kafri_16} (generalized eigenstate thermalization for integrable Hamiltonians~\cite{cassidy2011generalized, Vidmar_2016}) $\hat{H}_0$ that the predictions of $\hat{\rho}_{\text{DE}}(\tau)$ for few-body operators agree with those of the thermal (generalized Gibbs) ensemble \cite{dalessio_kafri_16, Vidmar_2016, essler_fagotti_16, caux_review_16}. We first focus on the case in which $\hat{H}_0$ is nonintegrable with no local conservation law. In this case, the inverse temperature $\beta(\tau)$ alone characterizes the thermal (grand canonical) ensemble at $\tau$, $\hat{\rho}_{\text{GE}}(\tau)=\exp[-\beta(\tau)\hat{H}_0]/\text{Tr}\{\exp[-\beta(\tau)\hat{H}_0]\}$, where $\beta(\tau)$ is determined by the condition $\text{Tr}[\hat{H}_0\hat{\rho}_{\text{GE}}(\tau)] = \text{Tr}[\hat{H}_0\hat{\rho}(\tau)]$. Only when $\beta(\tau)\ll 1$ is that one expects $\Gamma(\tau)$ to become independent of $\beta(\tau)$, and $E(\tau)$ to approach $E_\infty$ as a single exponential.

To illustrate this, in the main panel of Fig.~\ref{Fig_rate_beta} we plot $|e(\tau)|$ (normalized by its initial value $|e(0)|$) for various initial inverse temperatures $\beta_I\in[0.01,0.5]$. The normalized energies $e(\tau)/e(0)$ for $\beta_I=0.033$ and $0.01$ exhibit a nearly identical exponential decay (within the times at which the NLCE has converged) implying that $\Gamma$ is independent of $\beta_I$ [hence, of $\beta(\tau)$] when $\beta_I\lesssim 0.03$. For $\beta_I\gtrsim 0.2$, one can still use exponentials to fit $e(\tau)$, but the rates obtained depend on $\beta_I$. In the inset in Fig.~\ref{Fig_rate_beta}, we report the rates obtained from such fits vs $\beta_I$ using two orders of the NLCE and for two values of $g$. The rates from the two orders of the NLCE agree with each other and agree well with Fermi's golden rule predictions. (A worse agreement is seen for $g=0.5$ than for $g=0.2$ due to the effect of higher order corrections.) The increase in the rate seen in the inset in Fig.~\ref{Fig_rate_beta} with decreasing $\beta_I$ is the one expected to occur as a function of driving time for initial states that are not in the regime $\beta_I\ll 1$.

Next, we focus on the dependence of the heating rates on $\Omega$. In nonintegrable systems, the eigenstate thermalization hypothesis~\cite{deutsch1991quantum, srednicki1994chaos, rigol2008thermalization, dalessio_kafri_16} allows one to compute $\Gamma_m(\tau)$. After resolving all symmetries of the static Hamiltonian, the eigenstate thermalization hypothesis ansatz for the matrix elements $K^{(s)}_{i,f} = \bra{E^0_i}\hat{K}\ket{E^0_f}$ of the operator $\hat{K}$ (used as drive) in each block diagonal sector $s$ of $\hat{H}_0$ has the form \cite{dalessio_kafri_16, Srednicki_1999} 
\begin{eqnarray}\label{ETH_eq}
K^{(s)}_{i,f}= K^{(s)}(E)\delta_{i,f}+[D^{(s)}(E)]^{-1/2}f^{(s)}_K(E,\omega)R_{i,f},\ \
\end{eqnarray}
where $E=(E_i+E_f)/2$, $\omega=E_f-E_i$, $D^{(s)}(E)$ is the density of states of $\hat{H}_0$ in sector $s$ at energy $E$, and $R_{i,f}$ is a random variable with zero mean and unit variance. $K^{(s)}(E)$ and $f^{(s)}_K(E,\omega)$ are smooth functions of their arguments. 

Using Eqs.~\eqref{FGR_eq} and~\eqref{ETH_eq}, changing sums over eigenstates by integrals over energy, replacing $\hat{\rho}_{\text{DE}}(\tau)$ by $\hat{\rho}_{\text{GE}}(\tau)$ and assuming high temperature [$\beta(\tau)\ll1$], one obtains the following expression for the heating rate~\cite{suppmat}
\begin{eqnarray}
\Gamma_m=&&\dfrac{2\pi (m\Omega g_m)^2}{\text{Tr}(\hat{H}_0^2)} \sum_{s}\int_{E^{(s)}_{\text{min}}+m\Omega/2}^{E^{(s)}_{\text{max}}-m\Omega/2} dE\, |f_K^{(s)}(E,m\Omega)|^2\nonumber\\
&&\hspace{0.1cm}\times D^{(s)}(E+m\Omega/2) D^{(s)}(E-m\Omega/2)/D^{(s)}(E),\label{FGR_rate}
\end{eqnarray} 
where $E^{(s)}_{\text{min}}$ ($E^{(s)}_{\text{max}}$) is the minimum (maximum) energy in sector $s$, and $m\Omega$ is smaller than $E^{(s)}_{\text{max}}-E^{(s)}_{\text{min}}$ (otherwise there is no linear response heating for that mode).

\begin{figure*}[!t]
\includegraphics[width=0.99\textwidth]{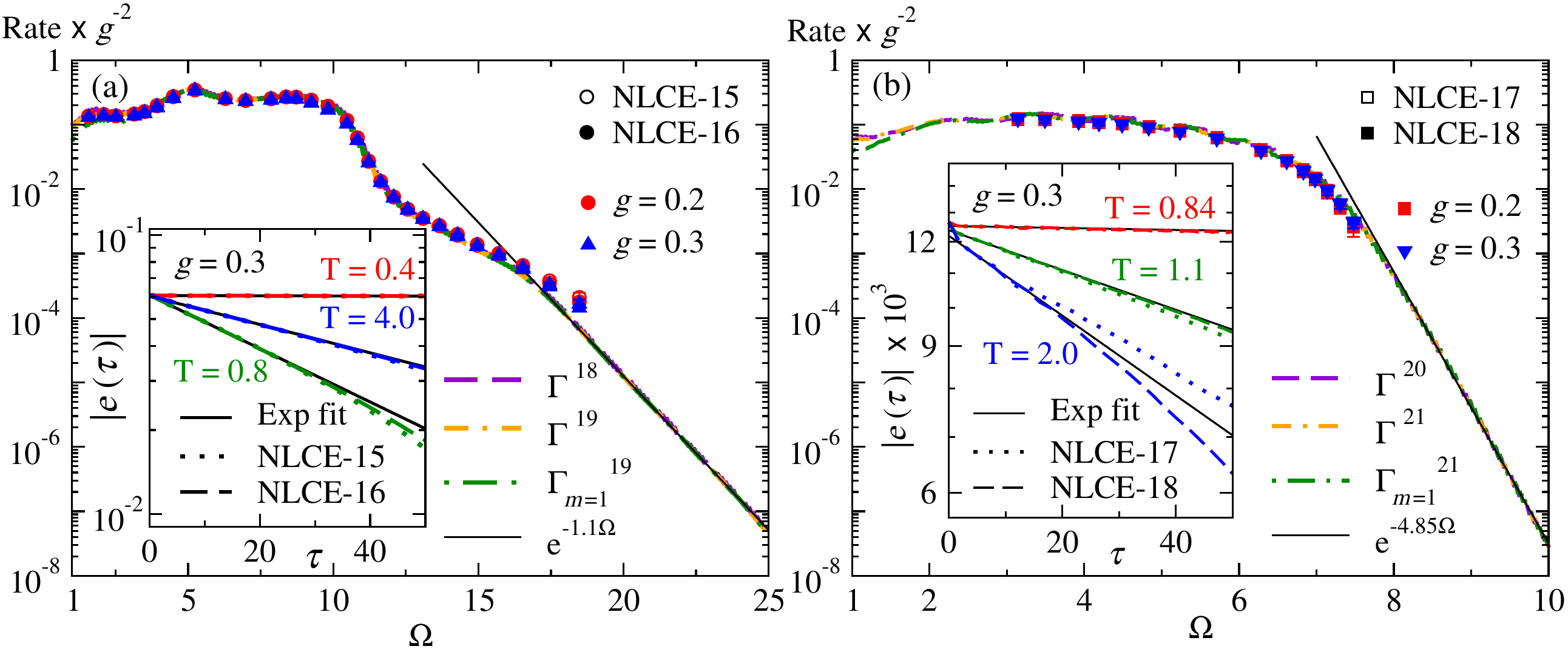}
\vspace{-0.1cm}
\caption{(Main panels) Heating rates (normalized by $g^2$) vs $\Omega$ for (a) the nonintegrable and (b) the integrable $\hat{H}_0$, for $g=0.2$ and $g=0.3$. Rates obtained from exponential fits of the dynamics (as in the insets) are shown as symbols for NLCE to (a) 15 (NLCE-15) and 16 (NLCE-16) orders, and (b) to 17 (NLCE-17) and 18 (NLCE-18) orders. Rates obtained from Eq.~(\ref{FGR_rate}) evaluated using full exact diagonalization in periodic chains are shown as lines for (a) 18 ($\Gamma^{18}$) and 19 ($\Gamma^{19}$) sites, and (b) 20 ($\Gamma^{20}$) and 21 ($\Gamma^{21}$) sites. We also show rates of the Fourier mode $m=1$ in Eq.~(\ref{FGR_rate}) for (a) 19 [$\Gamma_{m=1}^{19}$] and (b) 21 [$\Gamma_{m=1}^{21}$] sites, as well as exponential fits of the results at high $\Omega$. (Insets) Absolute value of the energy per site $|e(\tau)|$ vs $\tau$, using NLCE to (a) 15 (NLCE-15) and 16 (NLCE-16) orders and (b) 17 (NLCE-17) and 18 (NLCE-18) orders, for $g=0.3$ and three different driving periods $T=2\pi/\Omega$. Exponential fits to the highest order of the NLCE are shown as solid lines. The rates reported in the main panels are obtained from exponential fits for (a) $3\le\tau\le15$ for NLCE-16 and $3\le\tau\le12$ for NLCE-15, and (b) $2\le\tau\le8$ for NLCE-18 and $2\le\tau\le7.5$ for NLCE-17, for all $g$ and $T$ (error bars indicate fitting errors).}\label{Fig_rate_w}
\end{figure*}         

In Fig.~\ref{Fig_rate_w}(a), we compare heating rates (for the nonintegrable case and normalized by $g^2$) obtained from dynamics evaluated with NLCE (see inset) and the ones predicted by Eq.~(\ref{FGR_rate})~\cite{suppmat}. NLCE results are not reported for small and large values of $\Omega$ because the time interval in which the NLCE converges is not sufficiently long to produce robust exponential fits. The normalized rates for $g=0.2$ and $g=0.3$ are nearly identical to one another, and are well described by Eq.~(\ref{FGR_rate}). For high $\Omega$, we find that the evaluation of Eq.~(\ref{FGR_rate}) results in heating rates that can be well described by an exponential in $\Omega$. This is consistent with rigorous bounds~\cite{Abanin_2015_Exponentially, abanin2017rigorous, Else_2017_Prethermal}.

When $\hat{H}_0$ is integrable (the spin-1/2 $XXZ$ limit), the prethermal states are described by a generalized Gibbs ensemble (GGE) $\hat{\rho}_{\text{GGE}}(\tau)$~\cite{Wouters2014, Pozsgay2014, Ilievski2015}. When $\hat{\rho}_I$ is a thermal state with $\beta_I\ll1$ (or in general after long driving times), $\hat{\rho}_{\text{GGE}}(\tau)\simeq\hat{\rho}_{\text{GE}}(\tau)$ with $\beta(\tau)\ll1$~\cite{He_2012}. In this regime, Eq.~(\ref{FGR_rate}) gives the heating rates for the integrable static Hamiltonian provided that there is a well defined $|f_K^{(s)}(E,\omega)|^2$. In Fig.~\ref{Fig_rate_w}(b), we show the equivalent of Fig.~\ref{Fig_rate_w}(a) but for the integrable case. Despite the differences between the dependence of the heating rates on $\Omega$ in the nonintegrable and integrable cases, the heating rates in the latter are described by Eq.~(\ref{FGR_rate}) and, for high $\Omega$, they are well described by an exponential in $\Omega$.

The previous results show that heating rates can be used to probe the function $f_K^{(s)}(E,m\Omega)$ in nonintegrable and integrable systems. Still, Eq.~\eqref{ETH_eq} involves the density of states. For large system sizes, since $E$ is extensive but $\Omega$ is not, $D^{(s)}(E+ m\Omega/2) D^{(s)}(E- m\Omega/2) \simeq [D^{(s)}(E)]^2$ and $E^{(s)}_{\text{min,max}}\pm m\Omega/2 \simeq E^{(s)}_{\text{min,max}}$. Using the saddle point approximation to compute the integral in Eq.~\eqref{FGR_rate}, and using that $D^{(s)}(E_\infty)$ is maximal, the heating rate for Fourier mode $m$ in the thermodynamic limit $({\Gamma}^\infty_m)$ can be written as
\begin{eqnarray}
{\Gamma}^\infty_m=\dfrac{2\pi (m\Omega g_m)^2}{\text{Tr}(\hat{H}_0^2)} \sum_{s} |f_K^{(s)} (E_\infty,m\Omega)|^2 Z(s), \label{FGR_ME}
\end{eqnarray}  
where $Z(s)$ is the Hilbert space dimension of sector $s$. Thus, the rate for Fourier mode $m=1$, which Fig.~\ref{ETH_eq} shows to be in excellent agreement with the heating rates obtained from the NLCE dynamics for a wide range of values of $\Omega$, gives the average $|f_K^{(s)} (E_\infty,\Omega)|^2$ over all sectors of the Hamiltonian in the thermodynamic limit~\cite{suppmat}. 

In summary, we studied heating in strongly interacting driven lattice systems and showed that, at sufficiently high effective temperatures ($[\beta(\tau)]^{-1}\gtrsim 2$), it can be well characterized by rates no matter whether the system is nonintegrable or integrable. We also showed that the rates agree with Fermi's golden rule predictions for both nonintegrable or integrable cases. We then argued that  heating rates can be used to probe the structure of off-diagonal matrix elements of the operator used to drive the system, in the eigenstates of the static Hamiltonian. Our results suggest that there is a well defined $|f_K^{(s)} (E,\Omega)|^2$ in integrable interacting systems. This has been confirmed in a recent full exact diagonalization study of the spin-1/2 $XXZ$ chain~\cite{LeBlond2019}, and needs to be further explored to place it on equal footing with what is known for quantum chaotic systems~\cite{khatami_pupillo_13, steinigeweg_herbrych_13, beugeling_moessner_15, dalessio_kafri_16, luitz_17, mondaini_rigol_17, jansen_stolpp_19}.

\begin{acknowledgements}
This work was supported by the National Science Foundation under Grant No.~PHY-1707482. We are grateful to W. De Roeck and S. Gopalakrishnan for motivating discussions. The computations were carried out at the Institute for CyberScience at Penn State.
\end{acknowledgements}

\bibliographystyle{apsrev4-1}

\bibliography{Reference}

\newpage
\phantom{a}
\newpage
\setcounter{figure}{0}
\setcounter{equation}{0}
\setcounter{section}{0}

\renewcommand{\thetable}{S\arabic{table}}
\renewcommand{\thefigure}{S\arabic{figure}}
\renewcommand{\theequation}{S\arabic{equation}}
\renewcommand{\thesection}{S\arabic{section}}

\onecolumngrid

\begin{center}

{\large \bf Supplemental Material:\\
Heating rates in periodically driven strongly interacting quantum many-body systems
}\\

\vspace{0.3cm}

Krishnanand Mallayya and Marcos Rigol\\
{\it Department of Physics, The Pennsylvania State University, University Park, PA 16802, USA}

\end{center}

\onecolumngrid

\label{pagesupp}

\section{S1. Numerical evaluation of Eq.~(1) in the main text\label{sec_S1}}

Equation~(1) in the main text is evaluated using full exact diagonalization of chains with $L$ sites and periodic boundary conditions. Defining a small energy window $\Delta E$, Eq.~(1) is modified to the following expression (which is amenable to numerical evaluation)
\begin{eqnarray}\label{supp_fermi_DE}
\dot{E}_{\Delta E,m}(\tau) = \dfrac{2\pi g_m^2}{\Delta E}\sum_{i}P_i^0(\tau)\times\sum_{|E_f^0-E_i^0\pm m\Omega|\le\Delta E/2} \left|\bra{E^0_f}\hat{K}\ket{E^0_i}\right|^2\left(E_f^0-E_i^0\right),
\end{eqnarray}
where $\ket{E^0_{i}}$ $(\ket{E^0_f})$ are eigenkets of $\hat{H}_0$ with eigenenergies $E^0_{i}$ ($E^0_f$), and $P_i(\tau)=\bra{E^0_i}\hat{\rho}(\tau)\ket{E^0_i}$. With this coarse graining procedure, $\Gamma_{\Delta E,m}(\tau)$ for Fourier mode $m$ is calculated as 
\begin{eqnarray}
\Gamma_{\Delta E,m}(\tau)= \dfrac{\dot{E}_{\Delta E,m}(\tau)}{E_{\infty}-E(\tau)}\label{supp_Gamma_fermi},
\end{eqnarray}  
where $E(\tau)$ is also evaluated using full exact diagonalization, and $E_\infty=0$ for our model. $\Gamma_{\Delta E}(\tau)=\sum_{m>0}\Gamma_{\Delta E,m}(\tau)$ is the relaxation rate of $E(\tau)$.

\begin{figure*}[!b]
\includegraphics[width=0.98\textwidth]{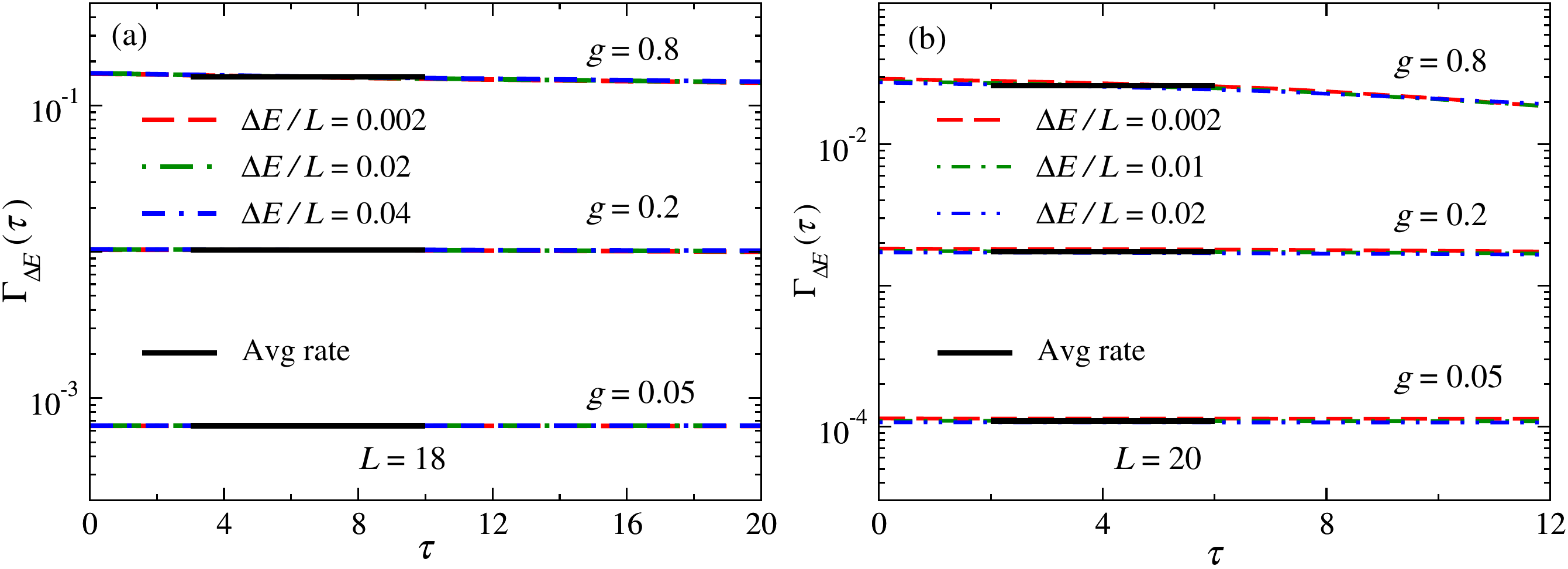}
\vspace{-0.1cm}
\caption{$\Gamma_{\Delta E}(\tau)=\sum_{m>0}\Gamma_{\Delta E,m}(\tau)$ [see Eq.~(\ref{supp_Gamma_fermi})] evaluated using full exact diagonalization of chains with $L$ sites and periodic boundary conditions for three values of $\Delta E$ for (a) the nonintegrable and (b) the integrable static Hamiltonians. Each system is driven with $g=0.05,\, 0.2,\, 0.8$, a period $T=1.0$, and the initial state is at an inverse temperature $\beta_I=(30)^{-1}$. The solid line is the average $\Gamma_{\Delta E}(\tau)$ over (a) $\Delta E/L=\{0.002,0.004,\dots,0.04\}$ and $\tau=\{3,4,\dots,10\}$ (a total of 160 values) for the nonintegrable $\hat{H}_0$, and (b) $\Delta E/L=\{0.002,0.004,\dots,0.02\}$ and $\tau=\{2,3,\dots,6\}$ (a total of 50 values) for the integrable $\hat{H}_0$.}  \label{Fig_supp_FGR}
\end{figure*}  

In Fig.~\ref{Fig_supp_FGR}, we show $\Gamma_{\Delta E}(\tau)$ vs $\tau$ for three values of $\Delta E$ when $g=0.05$, $0.2$, and $0.8$, for the nonintegrable [$L=18$, Fig.~\ref{Fig_supp_FGR}(a)] and the integrable [$L=20$, Fig.~\ref{Fig_supp_FGR}(b)] static Hamiltonians (the period of the drive is $T=1.0$). The initial thermal state has $\beta_I=(30)^{-1}$. It is apparent in Fig.~\ref{Fig_supp_FGR} that $\Gamma_{\Delta E}(\tau)$ is nearly constant, with a slight drift at long times (apparent for $g=0.8$), and that it is independent of the value of $\Delta E$. We identify a range of $\Delta E$ and $\tau$ where $\Gamma_{\Delta E}(\tau)$ is (nearly) constant [and where the dynamics of $e(\tau)$ is exponential and robust against finite-size effects, see Sec.~S3] and compute the average of $\Gamma_{\Delta E}(\tau)$ in this range.  

In the main text, all the rates reported in Figs.~1 and~2 for which Eq.~(1) was used were obtained averaging over $\Delta E/L=\{0.002,0.004,\dots,0.04\}$ and $\tau=\{3,4,\dots,10\}$ (a total of 160 values) for the nonintegrable $\hat{H}_0$, and $\Delta E/L=\{0.002,0.004,\dots,0.02\}$ and $\tau=\{2,3,\dots,6\}$ (a total of 50 values) for the integrable $\hat{H}_0$. The standard deviation of the averages were reported as error bars. 

\section{S2. Derivation of Eq.~(5) in the main text}

Equation~(1), accounting for the block diagonalization of $\hat{H}_0$ in symmetry sectors $\{s\}$, has the form
\begin{eqnarray}
\dot{E}_m(\tau)=2\pi g_m^2\sum_{s}\left(\sum_{i,f\in s}|K^{(s)}_{i,f}|^2(E^0_f-E^0_i)P_i^0(\tau)
\delta(E^0_f-E^0_i\pm m\Omega)\right)\label{supp_FGR1}
\end{eqnarray}  
where $K^{(s)}_{i,f}=\bra{E^0_i}\hat{K}\ket{E^0_f}$, for $i,f\in s$, and $P_i^0(\tau) = \bra{E^0_i}\hat{\rho}(\tau)\ket{E^0_i}$. From the eigenstate thermalization hypothesis (ETH) it follows that the results from the diagonal ensemble and the Gibbs ensemble agree~\cite{deutsch1991quantum, srednicki1994chaos, rigol2008thermalization, dalessio_kafri_16}, so one can replace $P_i^0(\tau)$ by $\exp[-\beta(\tau)E^0_i]/Z(\tau)$, where $Z(\tau)$ is the partition function, and the inverse temperature $\beta(\tau)$ is set by the energy $E(\tau)$. 

Using the Gibbs ensemble, the ETH ansatz for $K^{(s)}_{i,f}$ (see main text), and replacing sums by integrals, Eq.~\eqref{supp_FGR1} can be written as
\begin{eqnarray}
\dot{E}_m(\tau)=2\pi m\Omega (g_m)^2 \sum_{(s)}&&\left\{\int_{E^{(s)}_{\text{min}}}^{E^{(s)}_{\text{max}}-m\Omega} \dfrac{dE\, e^{-\beta(\tau)E}}{Z(\tau)} \left|f_K^{(s)}(E+m\Omega/2,m\Omega)\right|^2\dfrac{ D^{(s)}(E)D^{(s)}(E+m\Omega)}{D^{(s)}(E+m\Omega/2)}\right.\nonumber\\&& \left. -\int_{E^{(s)}_{\text{min}}+m\Omega}^{E^{(s)}_{\text{max}}} \dfrac{dE\, e^{-\beta(\tau)E}}{Z(\tau)} \left|f_K^{(s)}(E-m\Omega/2,m\Omega)\right|^2\dfrac{ D^{(s)}(E)D^{(s)}(E-m\Omega)}{D^{(s)}(E-m\Omega/2)} \right\},
\end{eqnarray} 
where $E^{(s)}_{\text{min}}$ ($E^{(s)}_{\text{max}}$) is the minimum (maximum) energy in sector $s$, and we used that $|f_K^{(s)}(E,-\omega)|=|f_K^{(s)}(E,\omega)|$. A change of variable $E\rightarrow E+m\Omega/2$ in the first integral, and $E\rightarrow E-m\Omega/2$ in the second integral, allows one to rewrite the expression above as
\begin{equation}
\dot{E}_m(\tau)=4\pi m\Omega (g_m)^2\sinh\left[\dfrac{\beta(\tau)m\Omega}{2}\right]
\sum_{s}\int_{E^{(s)}_{\text{min}}+m\Omega/2}^{E^{(s)}_{\text{max}}-m\Omega/2} \dfrac{dE\, e^{-\beta(\tau)E}}{Z(\tau)} \left|f_K^{(s)}(E,m\Omega)\right|^2 \dfrac{D^{(s)}(E+m\Omega/2) D^{(s)}(E-m\Omega/2)}{D^{(s)}(E)} .\label{supp_Em}
\end{equation}
At high temperatures, when $\beta(\tau)\ll1$, one has to lowest order in $\beta(\tau)$ 
\begin{equation}
\sinh\left[\dfrac{\beta(\tau)m\Omega}{2}\right]\simeq \dfrac{\beta(\tau)m\Omega}{2}, \quad e^{-\beta(\tau)E}\simeq 1, \quad Z(\tau)\simeq \text{Tr}[1], \quad \text{and} \quad [E_\infty-E(\tau)]\simeq\dfrac{\beta(\tau)\text{Tr}[\hat{H}_0^2]}{\text{Tr}[1]}.  \label{highT} 
\end{equation}
Using Eqs.~\eqref{supp_Em} and~\eqref{highT}, the heating rate $\Gamma_m=\dot{E}_m(\tau)/[E_\infty-E(\tau)]$ reduces to Eq.~(5) in the main text. 

\subsection{Numerical evaluation of Eq.~(5) in the main text}

Like Eq.~(1), Eq.~(5) in the main text is evaluated using full exact diagonalization of chains with $L$ sites and periodic boundary conditions. We define a small energy window $\Delta E$, which we use to bin the spectrum of $\hat{H}_0$ in each symmetry sector $s$. Each bin $\alpha$, with energy $E_\alpha$, includes all eigenstates with eigenenergies $E_i^0\in (E_\alpha-\Delta E/2,E_\alpha+\Delta E/2)$. The density of states at energy $E_\alpha$ is then $D^{(s)}(E_\alpha)=n_\alpha/\Delta E$, where $n_\alpha$ is the number of energy eigenstates in bin $\alpha$. The function $|f_K^{(s)}(E_\alpha,\omega_\alpha)|^2$, with $\omega_\alpha>0$, after coarse graining is given by
\begin{eqnarray}
\overline{|f_K^{(s)}(E_\alpha,\omega_\alpha)|^2}=D^{(s)}(E_\alpha)\left(\dfrac{\sum_{j,k}|\bra{E_j^0}\hat{K}\ket{E_k^0}|^2}{\sum_{j,k}1}\right)\label{supp_fEw},
\end{eqnarray}  
where $E^0_j$ and $E^0_k$ are such that bin $\alpha_j$ containing $E_j^0$ and $\alpha_k$ containing $E_k^0$ satisfy  $(E_{\alpha_j}+E_{\alpha_k})/2\in (E_\alpha-\Delta E/2, E_\alpha+\Delta E/2)$ and $|E_{\alpha_j}-E_{\alpha_k}|=\omega_\alpha$. This coarse graining procedure modifies Eq.~(5) in the main text to 
\begin{eqnarray}
\Gamma^{L}_{m}=\dfrac{2\pi (m\Omega g_m)^2}{\text{Tr}(\hat{H}_0^2)} \sum_{s}\sum_{\alpha} \Delta E\, \overline{|f_K^{(s)}(E_\alpha,m\Omega)|^2}\, \dfrac{D^{(s)}(E_\alpha+m\Omega/2) D^{(s)}(E_\alpha-m\Omega/2)}{D^{(s)}(E_\alpha)},\label{supp_Rcoarse}
\end{eqnarray} 
where the inner sum is over all the bins $\alpha$ whose energy $E_\alpha\in(E^{(s)}_{\text{min}}+m\Omega/2, E^{(s)}_{\text{max}}-m\Omega/2)$. 

In contrast to Eq.~(\ref{supp_fermi_DE}), Eq.~(\ref{supp_Rcoarse}) does not involve calculating the time evolution of the system. As a result, we are able to evaluate Eq.~(\ref{supp_Rcoarse}) in chains with $L=19$ ($L=21$) for the nonintegrable (integrable) static Hamiltonian. The dimension of the largest symmetry resolved sector is 13,797 (16,796) for the nonintegrable (integrable) $\hat{H}_0$.    

\begin{figure*}[!t]
\includegraphics[width=0.96\textwidth]{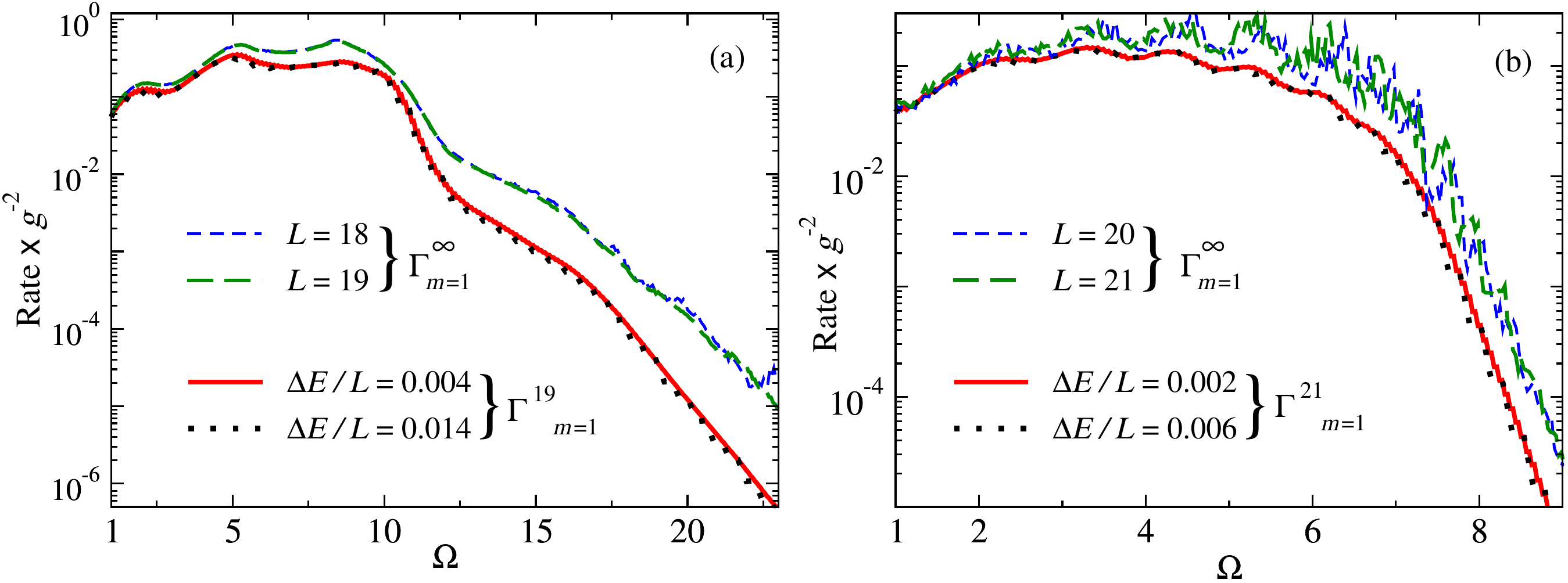}
\vspace{-0.1cm}
\caption{Rate (normalized by $g^{2}$) for the $m=1$ mode, $\Gamma^L_{m=1}$ [see Eq.~(\ref{supp_Rcoarse})] and $\Gamma^{\infty}_{m=1}$ [see Eq.~(\ref{supp_FGR_ME})], evaluated using full exact diagonalization of chains with $L$ sites and periodic boundary conditions for (a) the nonintegrable and (b) the integrable static Hamiltonians, for two values of $\Delta E$. The rates are evaluated at $\Omega=\Delta E, 2\Delta E, \dots$, for each value of $\Delta E$. For the spectrum of $\hat{H}_0$, the values: (a) $\Delta E/L\simeq 0.014$ and (b) $\Delta E/L\simeq 0.006$ correspond to $10L$ bins, and (a) $\Delta E/L\simeq 0.004$ and (b) $\Delta E/L\simeq 0.002$ correspond to $40L$ bins. For $\Gamma^{\infty}_{m=1}$, in both panels, results are reported for $\Delta E$ corresponding to $40L$ bins for the two largest chain sizes $L$ studied.} \label{Fig_supp_ratew}
\end{figure*}

In Fig.~\ref{Fig_supp_ratew}, we show heating rates for the $m=1$ mode, $\Gamma^{L}_{m=1}$, evaluated at $\Omega=\Delta E, 2\Delta E, \dots$ for two values of $\Delta E$ for the nonintegrable and the integrable static Hamiltonians. Our values of $\Delta E/L$ are such that the spectrum of $\hat{H}_0$ is divided into $10L$ bins [(a) $\Delta E/L\sim 0.014$ and (b) $\Delta E/L\sim 0.006$] and $40L$ bins [(a) $\Delta E/L\simeq 0.004$ and (b) $\Delta E/L\simeq 0.002$]. The results obtained can be seen to be robust against the choice of $\Delta E$. For the results reported in Fig.~\ref{Fig_rate_w} of the main text, we use $\Delta E/L\simeq 0.004$ ($40L$ bins) to evaluate $\Gamma^{18}$, $\Gamma^{19}$, and $\Gamma_{m=1}^{19}$ in Fig.~\ref{Fig_rate_w}(a) for the nonintegrable static Hamiltonian, and $\Delta E/L\simeq0.002$ ($40L$ bins) to evaluate $\Gamma^{20}$, $\Gamma^{21}$, and $\Gamma_{m=1}^{21}$ in Fig.~\ref{Fig_rate_w}(b) for the integrable static Hamiltonian. 

In Fig.~\ref{Fig_supp_ratew}, we also show results of the numerical evaluation of Eq.~(6) in the main text using full exact diagonalization of chains with periodic boundary conditions and $L$ sites. The coarse grained Eq.~(6), using Eq.~\eqref{supp_fEw}, has the form
\begin{eqnarray}
{\Gamma}^\infty_m=\dfrac{2\pi (m\Omega g_m)^2}{\text{Tr}(\hat{H}_0^2)} \sum_{s} \overline{|f_K^{(s)} (0,m\Omega)|^2} Z(s). \label{supp_FGR_ME}
\end{eqnarray} 
It is apparent in Fig.~\ref{Fig_supp_ratew}, both for the nonintegrable and the integrable static Hamiltonians, that the results for $\Gamma^{\infty}_{m=1}$ calculated using Eq.~\eqref{supp_FGR_ME} do not agree with the ones for $\Gamma^{L}_{m=1}$ using Eq.~\eqref{supp_Rcoarse}. This is because of strong finite-size effects in $\Gamma^{\infty}_{m=1}$. We note that the disagreement increases as $\Omega$ increases. The fact that finite-size effects in $\Gamma^{\infty}_{m=1}$ increase with increasing $\Omega$ is also apparent in the increasing discrepancy with increasing $\Omega$ between the results for the two chain sizes shown in Fig.~\ref{Fig_supp_ratew}. This is in contrast to the results for $\Gamma^{L}$ (similar to $\Gamma^{L}_{m=1}$ at large $\Omega$) evaluated from Eq.~\eqref{supp_Rcoarse} and reported in Fig.~3 in the main text for two systems sizes. The strong finite-size effects in $\Gamma^{\infty}_{m=1}$ are not surprising as the assumptions made to derive Eq.~(6) are not valid for the small system sizes studied in this work.

\section{S3. Convergence of NLCE and exact diagonalization}\label{sec:convergence}

\begin{figure}[!t]
\includegraphics[width=0.49\textwidth]{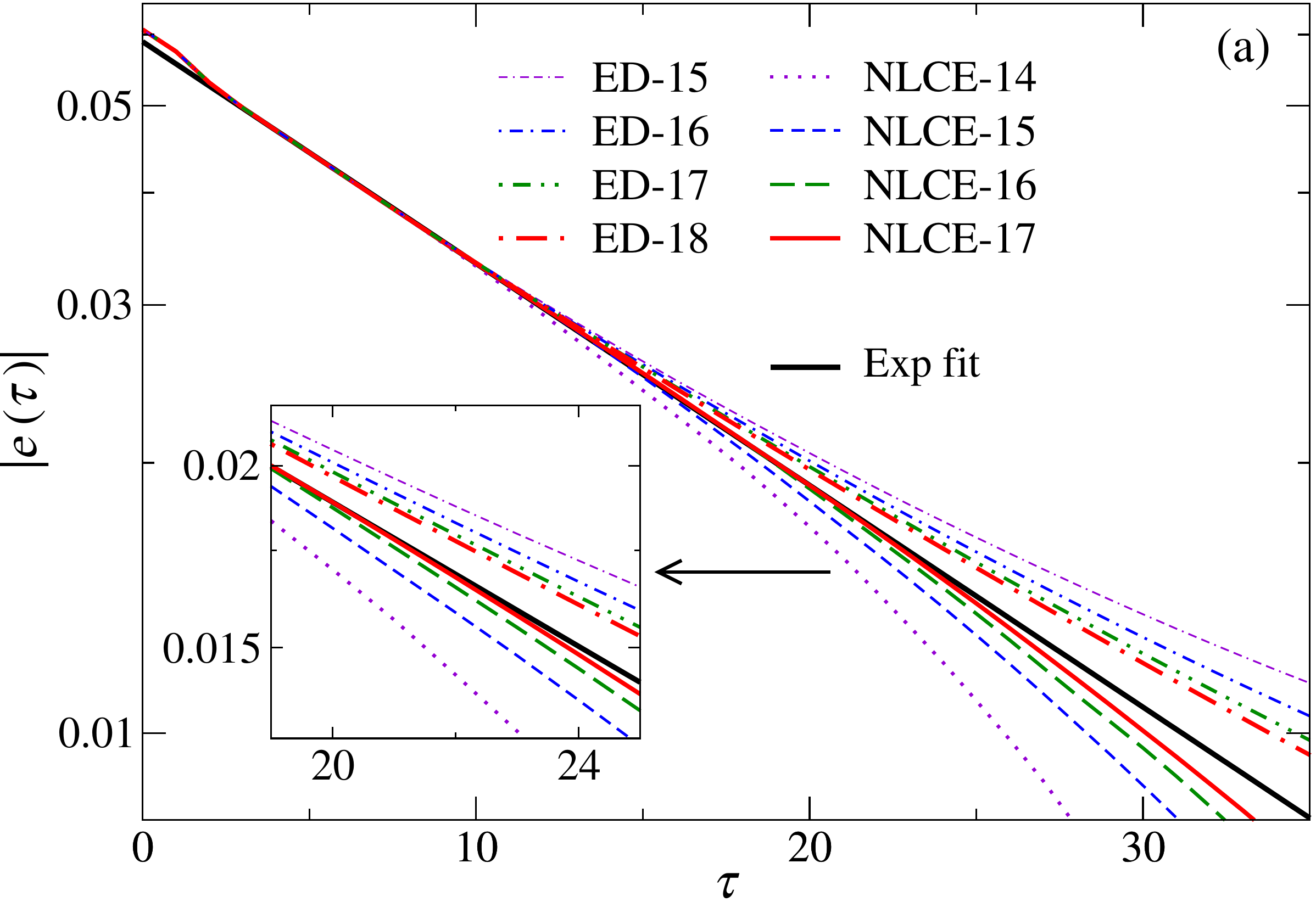}
\includegraphics[width=0.49\textwidth]{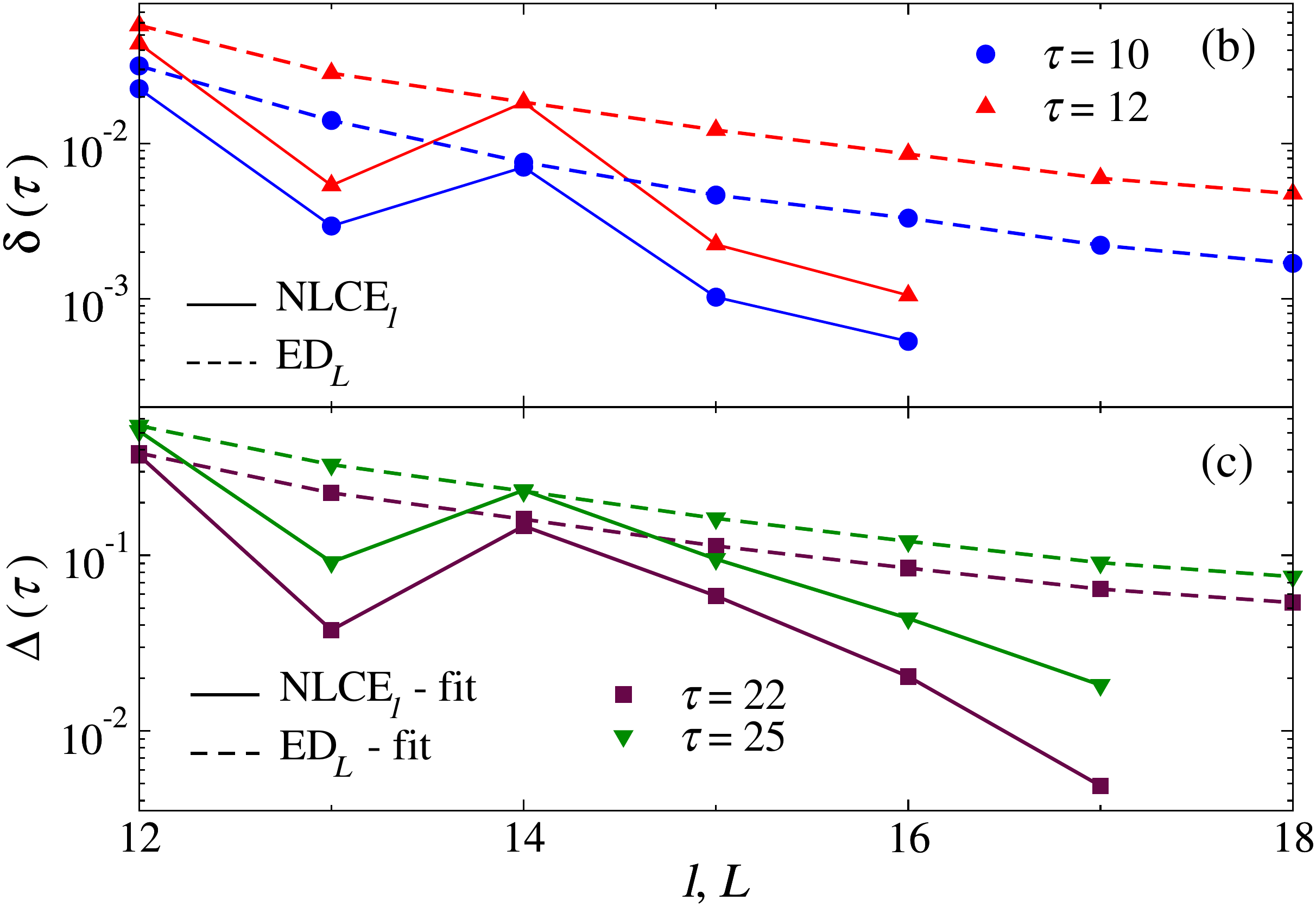}
\vspace{-0.1cm}
\caption{(a) Absolute value of the energy per site $|e(\tau)|$ of the nonintegrable $\hat{H}_0$ for $g=0.5$,  $T=1$, and $\beta_I=30^{-1}$, evaluated within the last four orders $l$ of the NLCE (NLCE-$l$) and for the four largest chain sizes $L$ calculated using ED (ED-$L$). An exponential fit to $|e(\tau)|$, using the ED-18 results for $3\le\tau\le10$, is also shown (Exp fit). Inset: zoom to the result in the main panel for $19\le\tau\le25$. (b) The errors $\delta^{l}_{\text{NLCE}}(\tau)$ and $\delta^{L}_{\text{ED}}(\tau)$ [Eq.~\eqref{supp_nlce_ed}] labeled as NLCE$_l$ and ED$_L$, respectively, at $\tau=10$ and $\tau=12$. (c) Normalized deviation from the exponential fit, $\Delta_l(\tau)$ and $\Delta_L(\tau)$ [Eq.~\eqref{supp_De_l_L}] for the $l^{\text{th}}$ NLCE order (NLCE$_l$-fit), and ED with $L$ sites (ED$_L$-fit), respectively, evaluated at $\tau=22$ and $\tau=25$. The exponential fit used is the one in panel (a).} \label{Fig_supp_NLCE_ED}
\end{figure}  

In Fig.~\ref{Fig_supp_NLCE_ED}(a), we plot the absolute value of the energy per site $|e(\tau)|$ of the nonintegrable Hamiltonian for $g=0.5$, period $T=1$, and $\beta_I=30^{-1}$. The results shown for $|e(\tau)|$ were obtained within the last four orders $l$ of the numerical linked cluster expansion (NLCE), and for the four largest chain sizes $L$ (with periodic boundary conditions) studied using exact diagonalization (ED). At short times ($\tau\lesssim10$), both NLCE and ED give nearly identical results (all lines are indistinguishable in the scale of the figure). The small finite-size effects of the ED calculations for $\tau\lesssim10$ are the reason we evaluate Eq.~(1) of the main text with ED in the range $3\le\tau\le10$ (see Fig.~\ref{Fig_supp_FGR}). All the curves in this range are well described by an exponential, so we fit an exponential to the results from the largest chain calculated with ED ($L=18$) in this range of $\tau$ [shown in Fig.~\ref{Fig_supp_NLCE_ED}(a) as a black line]. For $\tau\gtrsim10$, ED calculations deviate from the exponential, and with  increasing $L$ the curves monotonically approach the exponential fit. On the other hand, NLCE results ($l=16$ and $17$) remain exponential up to $\tau\sim20$ [apparent in the inset in Fig.~\ref{Fig_supp_NLCE_ED}(a)]. Increasing the order $l$ of the NLCE significantly improves the convergence towards the exponential at longer times. We remark here that, purely from the ED calculations, it is difficult to identify the exponential regime (in order to accurately predict the rates) as the results from ED smoothly drift away from an exponential and the discrepancies between $L=17$ and $18$ are small at most times. We use the NLCE results as reference in order to identify the time interval in which the ED results exhibit the ``correct exponential''. That time interval is then used to compute robust Fermi's golden rule predictions.

Next, we quantify the convergence errors of NLCE and the finite-size errors of ED calculations. At the times at which the highest order ($l=17$) NLCE results are very close to the exponential fits to the ED (and NLCE) data, the NLCE results serve best as reference to quantify the errors at lower orders of the NLCE and finite-size errors of ED. For $e(\tau)$ evaluated with the $l^{\text{th}}$ NLCE order [$e_l(\tau)$] and with the $L$-site periodic chain ED [$e_L(\tau)$], we define the convergence errors for $l<17$ and $L$, respectively, as the relative differences from $e_{l=17}(\tau)$ given by
\begin{eqnarray}\label{supp_nlce_ed}
\delta^l_{\text{NLCE}}(\tau)= \dfrac{|e_l(\tau)-e_{l=17}(\tau)|}{|e_{l=17}(\tau)|} \quad\text{and}\quad 
\delta^L_{\text{ED}}(\tau) = \dfrac{|e_L(\tau)-e_{l=17}(\tau)|}{|e_{l=17}(\tau)|}.
\end{eqnarray}

Fig.~\ref{Fig_supp_NLCE_ED}(b) reports $\delta^l_{\text{NLCE}}(\tau)$ and $\delta^L_{\text{ED}}(\tau)$ vs $l$ and $L$, respectively, at $\tau=10$ and $\tau=12$, for the same $e(\tau)$ as in Fig.~\ref{Fig_supp_NLCE_ED}(a). The plots make apparent that the errors decrease with increasing $l$ and $L$, that the errors at $\tau=12$ are greater than the corresponding ones at $\tau=10$, and suggest that the NLCE convergence errors decrease faster with increasing $l$ than the ED finite-size errors with increasing $L$, a known fact in equilibrium calculations~\cite{iyer_srednicki_15}. 

At $\tau=10$, the estimated of error for the ED calculation [$\delta^{L}_{\text{ED}}(10)$] for $L=18$ in Fig.~\ref{Fig_supp_NLCE_ED}(b) is less than 0.2\%. For all calculations with ED in this paper, $\tau=10$ is the largest time considered for the nonintegrable $\hat{H}_0$.

As argued in the main text, $e(\tau)$ in the thermodynamic limit is essentially a single exponential in $\tau$ for $\beta_I\lesssim30^{-1}$. Hence, for $\tau\gtrsim20$, we can estimate errors via the deviation of the NLCE and ED results from an exponential fit to the shorter-time results. To avoid any bias towards the NLCE results, here we compute the exponential fit from the ED results of $e_{L=18}(\tau)$ for $3\le\tau\le10$. The normalized deviation from the exponential fit for the $l^{\text{th}}$ NLCE order [$\Delta_l(\tau)$] and the $L$-site periodic chain solved with ED [$\Delta_L(\tau)$], are given by
\begin{eqnarray}\label{supp_De_l_L}
\Delta_l(\tau) = \dfrac{|e_l(\tau)-\text{fit}(\tau)|}{|\text{fit}(\tau)|} \quad\text{and}\quad \Delta_L(\tau) = \dfrac{|e_L(\tau)-\text{fit}(\tau)|}{|\text{fit}(\tau)|},
\end{eqnarray}
where fit($\tau$) is the value of the exponential fit at $\tau$. In Fig.~\ref{Fig_supp_NLCE_ED}(c),  $\Delta_l(\tau)$ and $\Delta_L(\tau)$ are plotted vs $l$ and $L$, respectively, for $\tau=22$ and $\tau=25$. It is apparent that the NLCE results converge faster towards the exponential [at $\tau=22$, $\Delta_{l=17}(\tau)$ is less than 0.5\%]. The errors in the ED calculations are an order of magnitude higher, and vanish more slowly with increasing $L$ [as in Fig.~\ref{Fig_supp_NLCE_ED}(b)]. The excellent convergence of the NLCE results (for $\tau\le20$ the nonintegrable $\hat{H}_0$) was essential for the accuracy of the relaxation rates computed via fits to the NLCE data that were reported in the main text.   

\end{document}